\documentclass[preprint,12pt]{elsarticle}

\usepackage{amssymb}
\usepackage{amsmath}

\usepackage{graphicx} 
\usepackage[position=bottom]{subfig}     
\usepackage{float}      
\usepackage{color}
\usepackage{soul}

\usepackage{xurl}                           

\usepackage[utf8]{inputenc}         
\usepackage[acronym,toc,nonumberlist]{glossaries-extra}    
\setkeys{glslink}{hyper=false}
\setabbreviationstyle[acronym]{long-short}          

\makeglossaries

\newglossaryentry{sup}
{
    name=superframe,
    description={Time is divided into contiguous superframes}
}

\newglossaryentry{ts}
{
    name=time slot,
    description={Each superframe is divided into time slots. A device can transmit during a time slot}
}

\newglossaryentry{g}
{
    name = $T_{G}$, 
    description={}
}

\newglossaryentry{gn}
{
    name = $T_{Gn}$,
    description={}
}

\newglossaryentry{s}
{
    name = $T_{S}$, 
    description={}
}

\newglossaryentry{sn}
{
    name = $T_{Sn}$, 
    description={}
}

\newglossaryentry{so}
{
    name = $T_{So}$, 
    description={}
}

\newglossaryentry{ct}
{
    name = CT, 
    description={}
}

\newglossaryentry{tsr}
{
    name = time slot retention,
    description={}
}

\newglossaryentry{dss}
{
    name = default superframe size,
    description={}
}

\newglossaryentry{fst}
{
    name = failed shrink timeout,
    description={}
}

\newglossaryentry{st}
{
    name = ST, 
    description={}
}

\newglossaryentry{gm}
{
    name = GM, 
    description={}
}

\newglossaryentry{step}
{
    name = STEP\_SIZE,
    description={}
}

\newglossaryentry{fs}
{
    name = \textit{failed\_shrink} cache,
    description={}
}

\newglossaryentry{lf}
{
    name = LeavingFlag,
    description={}
}

\newglossaryentry{rs}
{
    name = Resolved state,
    description={}
}

\newglossaryentry{ss}
{
    name = Start state,
    description={}
}

\newglossaryentry{as}
{
    name = Assignment state,
    description={}
}

\newglossaryentry{to}
{
    name = $T$,
    description={}
}

\newglossaryentry{wcr}
{
    name = worst-case receiver,
    description={}
}

\newglossaryentry{ttile}
{
    name = transmission tile,
    description={}
}

\newglossaryentry{atile}
{
    name = allocation tile,
    description={}
}

\newacronym{mac}{MAC}{Medium Access Control}
\newacronym{csma}{CSMA}{Carrier-Sense Multiple Access}
\newacronym{tdma}{TDMA}{Time Division Multiple Access}
\newacronym{u}{UAV}{Unmanned Aerial Vehicle}
\newacronym{snr}{SNR}{Signal to Noise Ratio}
\newacronym{sinr}{SINR}{Signal to Interference plus Noise Ratio}
\newacronym{osi}{OSI}{Open Systems Interconnection}
\newacronym{ieee}{IEEE}{Institute of Electrical and Electronics Engineers}
\newacronym{wlan}{WLAN}{Wireless Local Area Network}
\newacronym{qos}{QoS}{Quality of Service}
\newacronym{gps}{GPS}{Global Positioning System}
\newacronym{csma-ca}{CSMA/CA}{Carrier Sense Multiple Access with Collision Avoidance}
\newacronym{csma-cd}{CSMA/CD}{Carrier Sense Multiple Access with Collision Detection}
\newacronym{wifi}{Wi-Fi}{802.11 standards}
\newacronym{faa}{FAA}{Federal Aviation Authority}
\newacronym{vtol}{VTOL}{Vertical Take Off and Landing}
\newacronym{wsn}{WSN}{Wireless Sensor Network}
\newacronym{dar}{D-ART}{Distributed Assignment and Resolution of Time slots}
\newacronym{wave}{WAVE}{Wireless Access in Vehicular Environments}
\newacronym{pid}{PID}{Proportional-Integral-Derivative}
\newacronym{dst}{D-STR}{Distributed Self-allocated Time slot Reuse}
\newacronym{sst}{SST}{Sum of Squares Total}
\newacronym{sse}{SSE}{Sum of Squares Error}
\newacronym{ber}{BER}{Bit Error Rate}
\newacronym{per}{PER}{Packet Error Rate}
\newacronym{bpsk}{BPSK}{Binary Phase-Shift Keying}
\newacronym{gi}{GI}{Guard Interval}
\newacronym{vanet}{VANET}{Vehicular Ad-hoc Network}

\journal{Computer Communications}

\begin{document}

\begin{frontmatter}



\title{Distributed Self-allocated Time Slot Reuse: \\Multi-hop Communication in Rigid UAV Formations}


\author[1]{Amelia Samandari} 
\affiliation[1]{organization={Computer Science and Software Engineering Department, University of Canterbury}, addressline={amelia.samandari@canterbury.ac.nz}}

\author[2]{Andreas Willig}
\affiliation[2]{organization={Computer Science and Software Engineering Department, University of Canterbury}, addressline={andreas.willig@canterbury.ac.nz}}

\author[3]{Barry Wu}
\affiliation[3]{organization={Wireless Research Centre, University of Canterbury}, addressline={barry.wu@canterbury.ac.nz}}

\author[4]{Philippa Martin}
\affiliation[4]{organization={Electrical and Computer Engineering Department, University of Canterbury}, addressline={philippa.martin@canterbury.ac.nz}}

\begin{abstract}
Deployment of Unmanned Aerial Vehicles (UAVs) in autonomous formations necessitates accurate 
	and timely communication of safety information. A communication protocol that supports timely 
	and successful transfer of safety information between UAVs is therefore needed. This paper 
	presents \acrfull{dst}. Our D-STR protocol addresses the essential task of communicating 
	safety information in rigid \gls{u} formations with different network topologies, enabling 
	collision-free deployment of the formation. This is an important step for improving the safety 
	and practicality of \gls{u} formations in application scenarios that span a range of industries.
\end{abstract}



\begin{keyword}


Unmanned Aerial Vehicle (UAV) \sep Medium Access Control (MAC) 
\sep Time Division Multiple Access (TDMA) \sep Multi-hop communication
\sep UAV formation \sep Distributed TDMA \sep Time slot reuse 
\end{keyword}

\end{frontmatter}




\section{Introduction}
\label{sec:intro}

Multi-Unmanned Aerial Vehicle (UAV) systems have shown significant promise 
in a range of practical applications such as supporting agriculture, horticulture
and dairy practices, as well as transportation, disaster relief, and
infrastructure monitoring~\cite{shakeri2019design}. Multi-\gls{u}
systems have increased efficiency and fault tolerance compared to
deployment of a single
\gls{u}~\cite{say_inata_shimamoto}. Multi-\gls{u} systems also allow
for simultaneous observations from different angles, or using
different sensors~\cite{erdelj_natalizio}. The time required to
complete tasks is reduced when those tasks can be completed in
parallel by multiple \glspl{u}~\cite{kalyaev2017novel}. Further,
affordable pricing has allowed their penetration into varied
industries~\cite{bekmezci_sahingoz_temel}.

The focus of our work is on \glspl{u} in a rigid formation. The motion
of each \gls{u} in a formation is based on its physically-neighbored
\glspl{u}~\cite{Schranz2020Swarm, lakew2020routing, singh}. Rigid
formations are a formation type where the structure does not change
during the formation. Therefore, the \glspl{u} maintain the same
relative positions throughout deployment. Rigid topologies have been 
used for relay networks, product delivery, and monitoring gas 
plumes~\cite{silic2019field, mohamed_al-jaroodi_jawhar_idries_mohammed}.

To safely, effectively, and scalably complete a wide range of tasks, a
UAV formation deployment must be adaptable and free from physical
collisions~\cite{Skorobogatov2020Multiple}. The formation should
accomplish this without the need for intervention from human
controllers. Additionally, in many cases the mission should be able to
continue and the system complete its given task, or a subset of it,
even in the event of a given number of formation \gls{u}
failures~\cite{bekmezci_sahingoz_temel}. This requires communicating
safety information between formation
\glspl{u}~\cite{abdessameud_tayebi}.

The aim of this work is to design a protocol supporting collision-free
flight through the periodic transmission of safety beacons. As beacons
are transmitted periodically, a TDMA-type MAC protocol can be used to
support the successful reception of these beacons. The final
superframe size established by a TDMA-type protocol must be small
enough to satisfy any application requirements for uncertainty about
neighbor positions. Formation safety, or lack of physical collisions,
is more important than an `optimal' allocation. A strictly optimal
solution of a minimum superframe size is often not practically
achievable, so a modest number of excess time slots in the final
allocation are acceptable. A large number of excess slots is
undesirable as it would increase the update delay, which increases a
\gls{u}'s uncertainty about its neighbors. Therefore, the objective is
to reduce the time until the final allocation is reached while still
maintaining a good frequency of updates throughout formation flight.

This work presents \acrfull{dst}, a TDMA-type \gls{mac} protocol for
multi-hop \gls{u} formations that supports spatial reuse with
distributed \gls{u} time slot allocation.  A key aspect of \gls{dst}
is the ability to dynamically adjust the superframe size.

TDMA algorithms typically focus on either single-hop or multi-hop
scheduling~\cite{ergen2010tdma}. In single-hop scheduling,
transmitters are one hop from the receiver. Therefore, only one device
is permitted to transmit in a given slot. Multi-hop TDMA scheduling is
inherently more complex than single-hop due to the potential for
spatial reuse, where multiple nodes can transmit without conflict in
the same time slot if their receivers are in distinct parts of the
network~\cite{ergen2010tdma}.

Results using our D-STR protocol demonstrate that our protocol always
converges, maintains low overhead, and achieves good reuse, with a
$3.56$ times increase in the number of formation UAVs (from $5677$ to
$19927$) only incurring a $1.44$ times increase in the superframe 
length (from $107.7$ to $154.9$ time slots). In addition, all periodic 
UAV transmissions are successfully received in $\sim 17$
rounds for all considered numbers of UAVs in the formation.

This paper is structured as follows: Section~\ref{sec:related-work}
discusses related work, Section~\ref{sec:system-model} outlines our
system model, Section~\ref{sec:protocol-design} presents our D-STR
protocol, Section~\ref{sec:results} reports and discusses our results,
and Section~\ref{sec:conclusion} outlines possible future work and 
concludes our paper.

\section{Related Work}
\label{sec:related-work}

MAC protocols are needed for UAV formations because the transmission
medium is shared between all UAVs in the formation.  If no access
controls are put in place there is a possibility of packet collisions
due to simultaneous transmission. When packet collisions occur, the
integrity of the transmitted packets’ contents is lost and the UAVs do
not receive information about their neighbors. Therefore, a UAV may
have an increased uncertainty of the position and behavior of its neighbors,
particularly if multiple packets collide in succession. This increases
the risk of physical collisions.  These packet collisions can be
mitigated through the use of a MAC protocol.

For the purpose of collision-free flights, MAC protocols for UAV
formations should ideally offer reliable delivery of periodically
transmitted safety packets from a transmitting UAV to all other UAVs
within a given neighbourhood. Because the
structure of a formation is fixed and the relative positions of UAVs
do not change, particularly in the case of rigid formations, a
schedule should be established such that the delay between receiving
periodic safety packets is bounded and reduced.

While extensive research has considered routing protocols for
multi-UAV systems~\cite{lakew2020routing}, to the best
of our knowledge there are limited \gls{mac} protocols
proposed for the case of multi-\gls{u} systems, and no \gls{mac}
protocols proposed for distributed formation control through the
exchange of safety messages between \glspl{u}.

Some MAC protocols proposed for multi-UAV systems are
\gls{csma}-based~\cite{alshbatat_dong, temel2015lodmac,
ho_grotli_johansen, ho_grotli_shimamoto_johansen,
cooperative_relay_mac}. These protocols react to UAVs that need to
share safety information, rather than proactively addressing the
underlying need to periodically transmit safety
information. \gls{csma} has some advantages in comparison to
\gls{tdma}. This includes its relative simplicity, with no need for
time synchronization (which, however, on UAVs often
comes for free as they are equipped with a GPS receiver). It is also
more adaptable than \gls{tdma}, because there is no establishment of a
schedule, and therefore suits dynamic topologies.
        
However, it has been shown that TDMA performs better than CSMA in
scenarios where devices generate regular periodic
traffic~\cite{wang_dong_wang_jiang}. CSMA performs best when the
overall network traffic is low or where network traffic is limited to
a few nodes, reducing overall contention for medium access. When
network traffic is distributed across many nodes in the network, the
performance of CSMA degrades significantly and TDMA becomes a more
appropriate choice as it performs better in high-contention scenarios
in static networks with homogeneous traffic
distribution, such as the proposed use case.

Hybrid MAC protocols have been designed for the \gls{u}-assisted
\gls{wsn} use case, to capitalize on the strengths of both CSMA and
TDMA for formation scenarios where network traffic generation varies
across time~\cite{wang_dong_wang_jiang,song,poudel2020energy,yang2019medium,araghizadeh2016efficient}. 
These protocols incur additional overhead from managing both CSMA and 
TDMA communication. Additionally, some hybrid MAC protocols focus on 
the interaction between a single UAV and multiple sensor nodes, rather
than communication between multiple UAVs~\cite{Poudel2019Medium}.

Distributed TDMA protocols take advantage of the benefits of a regular
transmission schedule for the network, while eliminating the need for
a central controller to assign devices to time slots. Previous works
that propose distributed TDMA protocols, for use cases other than
\gls{u} formation deployment,
include~\cite{adhoc_mac,Rhee2009DRAND,desync_mac,dt_scs_mac}. 
When compared with centralized TDMA protocols, distributed TDMA protocols
provide higher reliability because they respond to changes faster and 
do not suffer from single point of failure~\cite{sgora2015survey}. However, 
distributed TDMA protocols have considerably higher complexity~\cite{wang2007deterministic}.

ADHOC MAC \cite{adhoc_mac} is a TDMA-based protocol that dynamically
establishes a reliable single-hop broadcast channel and is designed
for the case where the nodes have no limitations on their power
consumption and with the assumption that time synchronization is
manged outside the protocol's scope. The number of devices in the
two-hop neighbourhood must not be greater than the number of slots 
in a frame, which limits the density that can be supported by the 
frame size. Our proposed \gls{dst} protocol does not have the limitation 
of a fixed superframe size, and the superframe size can grow or shrink 
based on the number of nodes and their density.

Distributed Randomized TDMA Scheduling (DRAND)~\cite{Rhee2009DRAND}
and its subsequent variants L-DRAND++~\cite{sato2014power},
E-T-DRAND~\cite{li2017distributed_etdrand}, and
ET-BT-DRAND~\cite{li2017distributed_etbtdrand} have been proposed for
\glspl{wsn}. DRAND exchanges \textit{request}/\textit{grant} messages
between one-hop neighbors to determine a random order coloring that
translates to a collision-free allocation. The limitations of the
mechanism adopted by DRAND include the potentially long delay before a
given device is allocated to a time slot, particularly given that a
device will not send a request in every round and this request may not
be granted.  Furthermore, many control messages are sent between
devices, increasing overhead, but there are no guarantees that these
control messages will be received.  L-DRAND++~\cite{sato2014power},
E-T-DRAND~\cite{li2017distributed_etdrand}, and
ET-BT-DRAND~\cite{li2017distributed_etbtdrand} refine DRAND by
introducing priority in a neighborhood address contention in each
round, but do not appear to have solved the issue of neighbors
reliably receiving messages~\cite{nguyen2020distributed}.

Our proposed \gls{dst} protocol aims to reduce the delay before a
device is allocated to a time slot. The protocol requires separate
control messages to establish an allocation. Furthermore, the
successful reception of a transmission can be determined, and will not
require transmission of separate acknowledgment packets.

\section{System Model}
\label{sec:system-model}

In any \gls{u} formation, the \glspl{u} need to periodically
communicate with their neighbors for collision avoidance as well as
command and control. In this work, these periodic transmissions are
called \textbf{beacons}. A \gls{tdma}-type protocol is a suitable
choice for a \gls{u} formation, where the relative positions of the
\glspl{u} change very infrequently, if at
all~\cite{wang_dong_wang_jiang}. Because of the less dynamic nature of
a \gls{u} formation, once a transmission schedule is established it is
unlikely to change, or will only change after a significant period of
time or coming close to another formation working on the same
channel. Therefore, the initial overhead of using a \gls{tdma}-type
protocol amortizes over time~\cite{Rhee2009DRAND}. \Gls{tdma}-based
protocols have the additional benefit that packet collisions can be
avoided, and data rate and time guarantees can be
given~\cite{hadded2015tdma}. It should be noted that TDMA does not
perform carrier sensing and therefore is generally not as robust to
interference as CSMA~\cite{kurose2021networking}.

This work adopts a \gls{u} formation use case, so the relative
positions of the \glspl{u} are not expected to change throughout
deployment. The formation is assumed to be deployed in an area
containing no external physical obstacles. We assume that the
\glspl{u} are in their target positions from the outset.  The
\glspl{u} remain in the same relative position in the formation,
maintaining a given distance from the other \glspl{u} throughout
deployment. Therefore, the neighborhood of each \gls{u} does not
change.

In a \gls{u} formation that uses online control, where the UAVs make
decisions about their motion during flight, \glspl{u} need to
communicate periodically with their neighbors for collision avoidance
and command and control. Safety information is sent in a \textbf{beacon}. 
Beacons contain additional fields related to the D-STR protocol. These are 
discussed in detail in Section~\ref{sec:protocol-design}.

Each \gls{u} has a neighborhood of \glspl{u} within a \textbf{safety
radius} that it expects to receive \textbf{safety information} from.
The safety radius is a radius surrounding each \gls{u}.  Because the
use case is a rigid formation, we consider only the \glspl{u} within
the safety radius as relevant for collision avoidance.  A \gls{u}
within this safety radius is referred to as a \textbf{neighbor}.
    
Communication in \gls{u} formations features unique characteristics
(as compared to traditional wireless communications), including the
existence of distinct air-air and air-ground communication
channels~\cite{gutpa_survey_of_important_issues_uav}. Because our
focus is supporting the timely reception of this safety information,
the communication links that are of interest to this work are air-air
communications between \glspl{u} in the formation. It is also of note
that, within the context of this work, \textbf{collisions} will now
refer to transmission or packet collisions rather than physical
collisions.
    
Consider that formation \glspl{u} communicate over multiple air-air
channels. One of these channels is a dedicated `safety message
channel' that is only used to transmit beacons. Therefore, this method
does not preclude application data also being sent as this will be
done on a separate channel. This mirrors what is seen in the
international standard for communication between autonomous vehicles,
which also utilizes a dedicated channel for safety
messages~\cite{IEEEComputerSociety.LAN/MANStandardsCommittee.2010IEEEEnvironments}. 
Transmissions will use \gls{wifi} for the physical
layer~\cite{jawhar2017communication, lakew2020routing}. To allow for
more transmissions per unit of time, the size of these beacons should
be kept as small as possible. Therefore, only the most critical safety
information should be included in a beacon. It is assumed that the
underlying wireless technology (i.e.\ \gls{wifi}) has a packet checksum
and will discard packets with an incorrect checksum. This checksum is
assumed to be perfect. The formation is deployed high enough to
allow us to disregard reflections and
interference from the ground, and the same thermal
noise power is experienced by all formation \glspl{u}. All formation
\glspl{u} use the same transmit power and there is omnidirectional
propagation, with no reflections or shadowing, and therefore no
multi-path propagation. Path loss is considered.

A challenge with \gls{tdma} is agreement on the phase (i.e.\ where a
superframe starts) and the number of time slots in the superframe. The
time slots used in this work are long enough to support beacon
transmissions. This challenge is addressed by ensuring that superframe
information (i.e.\ the current time slot and superframe size) is
transmitted in each beacon, along with the safety information.

An additional challenge is synchronization of transmission time slots,
because the synchronization of the nodes' oscillators cannot be relied
upon over subsequent
transmissions~\cite{Nuspl1977Synchronization}. Using \gls{tdma} for a
formation of high-speed \glspl{u} is particularly susceptible to this
`clock drifting' and must be reliably
mitigated~\cite{Shrestha2018Precise, Pinto2018Robust}. Formation
\glspl{u} are fitted with \gls{gps} receivers. \Gls{gps} receivers
give highly precise timing information~\footnote{Time keeping using
\gls{gps} is theoretically accurate to approximately 14
nanoseconds~\cite{gps_time1, gps_time2}.} and can be used to manage
time synchronization~\cite{Hasan2018GNSS}.  Because the accuracy
afforded by GPS receivers is enough for the use case considered, the
issue of time synchronization is considered to be solved.

A further challenge of \gls{tdma} is optimizing the performance of the
network in cases where the traffic is unbalanced. This is relevant
when considering the transmission of application-specific data across
the formation and is not an issue when considering the transmission of
safety data in a formation. Every \gls{u} periodically transmits
safety data and therefore traffic is balanced.

This work is not concerned with the energy efficiency of the proposed
protocol designs. The reasons for this are twofold. Firstly, UAVs are
considered to have adequate energy for transmissions because the most
significant energy demand for any \gls{u} is from the propulsion
system~\cite{Adoni2023Investigation, lakew2020routing}. Secondly, \gls{u}
formations are delay-sensitive with respect to the timely transmission
of safety information to prevent physical collisions between
\glspl{u}~\cite{jawhar2017communication,Abdelkader2021Aerial,lakew2020routing}.
Therefore, energy requirements of the \gls{mac} protocol are secondary 
to its performance, and this work does not make design choices that favor
energy efficiency.

Our key performance metric in this work is the (average) time until
\textbf{convergence}. Convergence indicates the following condition:
all formation UAVs are allocated to a unique time slot, there are no
globally unallocated time slots, and each UAV's beacon transmissions
can be received by all neighboured UAV's with a sufficiently high
signal-to-interference-plus-noise ratio. This dynamic allocation does
not guarantee optimal time slot utilization, but does allow for
changes in the formation and fast convergence. True optimality (where
the superframe is the shortest possible length) is likely hard to
achieve due to the underlying optimization problem being NP-hard, and
time to convergence is more important for this use case than optimal
allocation~\cite{commander2009combinatorial}.

\section{Protocol Design}
\label{sec:protocol-design}

In the \gls{dst} protocol, which is based on \gls{tdma}, time is split
into adjoining \textbf{superframes}, that are in turn split into
adjoining time slots. The superframe consists of $N$ time slots. There
are 5 management time slots: Grow $T_{G}$, GrowNACK $T_{Gn}$, Shrink
$T_{S}$, ShrinkOBJECT $T_{So}$, and ShrinkNACK $T_{Sn}$. These are
followed by a number of transmission time slots, $T_{i}$ where
$i \in \{1,\ldots,N-5\}$, as seen in
Table~\ref{tab:superframe_structure}.  Transmission time slots are
used to transmit periodic beacons to update neighbors and management
time slots are used to modify the superframe size. The number of
required time slots is unknown and must be determined by the protocol.

\begin{table}[h]
    \centering
    \begin{tabular}{|l||c|c|c|c|c|c|c|c|c|c|}\hline
         & \multicolumn{10}{c|}{Superframe}\\\hline\hline
         Time Slot & $T_{G}$ & $T_{GN}$ & $T_{S}$ & $T_{SO}$    & $T_{SN}$ & $T_{1}$ & $T_{2}$ & $T_{3}$ & $T_{4}$ & $T_{5}$\\\hline
    \end{tabular}
    \caption{The superframe structure when $N$=10.} 
    \label{tab:superframe_structure}
\end{table}

To prevent physical collisions, the \gls{u} must know the position of
all its \textbf{neighbors}. A formation \gls{u} periodically sends a 
\textbf{beacon} to its neighborhood, containing safety data (e.g.\ the 
\gls{u}'s position, heading, speed), superframe data (e.g.\ the 
superframe size and current time slot that the transmission was 
sent/received during), and transmission success data (e.g.\ a record).

An outline beacon structure can be found in
Table~\ref{tab:dst_beacon_structure}. The superframe size and current
time slot fields provide superframe information and are used for
synchronization. The grow margin flag can be set to indicate that the
superframe should grow by \gls{gm} time slots. The time slot to remove
field nominates a time slot to eliminate from the superframe. The
leaving flag is set when a \gls{u} is planning to leave the
formation. The UAV ID can be set as the \gls{u}'s MAC address, or can
be set before deployment to another value that is unique
formation-wide. The safety information fields will, for example,
include position, heading, speed, and UAV size. The record is the last
field in the beacon, because it's size depends on the \gls{u}'s
neighborhood and therefore is variable.

\begin{table}[h]
	\centering
        \resizebox{\columnwidth}{!}{
	\begin{tabular}{|cccccccc|}
		\hline
		\multicolumn{8}{|c|}{Beacon}  \\ \hline \hline
		\multicolumn{1}{|c|}{\begin{tabular}[c]{@{}c@{}}Superframe\\ Size\end{tabular}} & \multicolumn{1}{c|}{\begin{tabular}[c]{@{}c@{}}Current\\ Slot\end{tabular}} & \multicolumn{1}{c|}{\begin{tabular}[c]{@{}c@{}}Grow\\ Margin\\ Flag\end{tabular}} & \multicolumn{1}{c|}{\begin{tabular}[c]{@{}c@{}}Slot to\\ Remove\end{tabular}} & \multicolumn{1}{c|}{\begin{tabular}[c]{@{}c@{}}Leaving\\ Flag\end{tabular}} & \multicolumn{1}{c|}{\begin{tabular}[c]{@{}c@{}}UAV \\ ID\end{tabular}} & \multicolumn{1}{c|}{\begin{tabular}[c]{@{}c@{}}Safety\\ Information\end{tabular}} & \begin{tabular}[c]{@{}c@{}}Record\end{tabular} \\ \hline
	\end{tabular}}
	\caption{An outline beacon structure.}
	\label{tab:dst_beacon_structure}
\end{table}

A \gls{u} may transmit a beacon during the \textbf{\gls{g}} or
\textbf{\gls{gn}} time slots to grow the superframe by appending
transmission time slots.  The \gls{g} slot is used to transmit a
request to grow the superframe. The \gls{gn} slot is used to indicate
unsuccessful reception (i.e.\ NACK) of a transmission in the \gls{g}
slot.
 
The \gls{s} slot is used to transmit a request to remove a given time
slot from the superframe. The \gls{so} and \gls{sn} time slots are
used to object to removal of the proposed time slot or to NACK the
transmission in the \gls{s} slot, respectively. To negotiate removal
of a time slot $T_{i}$, a \gls{u} may transmit a beacon during the
\textbf{\gls{s}}, \textbf{\gls{so}}, or \textbf{\gls{sn}} time slots.

Each time slot in the record can be set to one of three \textbf{settings}: 
`$0$' nothing received, `$1$' successfully received, and `$2$' decoding
failure. Successfully received indicates that a valid packet was
received, decoding failure indicates that energy was sensed in the
time slot but could not be decoded, and nothing received indicates
neither a valid packet nor a decoding failure. Each \gls{u} keeps
track of the setting it experiences for each time slot and updates its
record accordingly, as seen in Table~\ref{tab:beacon_record}. This
record is included in the beacon so the \glspl{u} can learn if their
transmission was successfully received.

\begin{table}[h]
    \centering
    \begin{tabular}{|l||c|c|c|c|c|c|c|c|c|c|}\hline
         Time Slot & $T_{1}$ & $T_{2}$ & $T_{3}$ & $T_{4}$ & $T_{5}$\\\hline
         Setting   & $1$     & $0$     & $2$     & $1$     & $1$\\\hline
         Record    & $01$    & $00$    & $10$    & $01$    & $01$ \\\hline
    \end{tabular}
    \caption{An example record maintained by a UAV. This is sent as a field in the beacon, where the settings are mapped to their binary representation in the record. For example, setting 2 is represented as 10 in the record.} 
    \label{tab:beacon_record}
\end{table}

The following protocol settings are known by all \glspl{u} and are
configured prior to deployment:
\begin{enumerate}
\item \textbf{The beacon structure.} Each beacon must contain safety
  data, superframe data and transmission success data.

\item \textbf{The time slot duration.} The time slot duration is
  sufficient to support beacon propagation. Because the beacon length
  is variable, the duration of a single time slot should be long
  enough to accommodate the largest beacon. The time slot duration
  stipulates an upper bound on the size of any neighborhood.
    
\item \textbf{The beacon transmit power setting.} This is used to send
  periodic beacons in a transmission time slot and is set such that
  \begin{enumerate}
  \item all neighbors can successfully decode the transmission, and
  \item there is a margin to account for the interference that could
    be generated by other \glspl{u} transmitting in the same time
    slot.
  \end{enumerate}

  That is to say, allowing multiple transmissions in the same time
  slot introduces interference. Therefore, reception success hinges on
  an acceptable \gls{sinr} value, rather than an acceptable \gls{snr}
  value, and the chosen transmit power must take this into account.

\item \textbf{The adapt transmit power setting.} This is used to send
  transmissions that must be received by the entire formation. That is
  to say, transmissions in the management time slots: \gls{g},
  \gls{gn}, \gls{s}, \gls{sn}, \gls{so}. This parameter is set based
  on the maximum geographic size the formation deployment can
  have. Therefore, the higher transmit power is set such that a
  transmission from any \gls{u} in the formation can reach all other
  \glspl{u} in the formation. Naturally, this will limit the
  geographical spread of the formation, since all \glspl{u} must be
  within the maximum range of the adapt transmit power.

\item \textbf{The \gls{ct}} (collision threshold).
  This indicates the maximum number of consecutive unsuccessful
  allocation attempts before a \gls{u} transmits in the grow time
  slot. Each \gls{u} keeps a count, $c$, of its number of consecutive
  failed self-allocation attempts. If this threshold is reached and
  $c=\gls{ct}$, the \gls{u} will transmit in the \gls{g} slot, new
  time slots will be added to the end of the superframe, and $c$ will
  be reset to zero.

\item \textbf{The superframe growth \gls{step}.} This specifies the
  number of time slots to add to the end of the superframe if multiple
  \glspl{u} transmit in the grow time slot, or the transmitting
  \gls{u} has reached \gls{ct} unsuccessful self-allocations. Note
  that a \gls{u} may also choose to grow the \gls{sup} if there are no
  available time slots.

\item \textbf{The \gls{st}} (shrinking
    threshold). This indicates the number of superframes that a
  \gls{u} observes a time slot's setting to be `0' before it transmits
  in the \gls{s}. Each \gls{u} keeps a count, $s$, of the number of
  superframes that it observes a time slot setting to be `0'. If this
  threshold is reached and $s$=\gls{st}, the \gls{u} will transmit in
  the \gls{s}.

    \item \textbf{The default starting superframe size.} If a \gls{u} is the first \gls{u} to `join' the formation, it will use the default superframe size and self-allocate to the transmission time slot. It is possible for an operator to set the default superframe size at the start of a mission, which can be chosen based on the size of the formation. The first \gls{u} to `join' the formation will self-allocate to the first transmission time slot.

    \item \textbf{The \gls{tsr} probability.} Once a \gls{u} has successfully self-allocated, it will continue to transmit in the same time slot unless it learns that its beacons are not being successfully received. Each successfully self-allocated \gls{u} will release its time slot with a probability $p$ if its transmission was not successfully received. 
    
\end{enumerate}

\subsection{UAV States}
\label{sec:dst_uav_states}

\begin{figure}
	\centering
	\includegraphics[width=0.8\columnwidth]{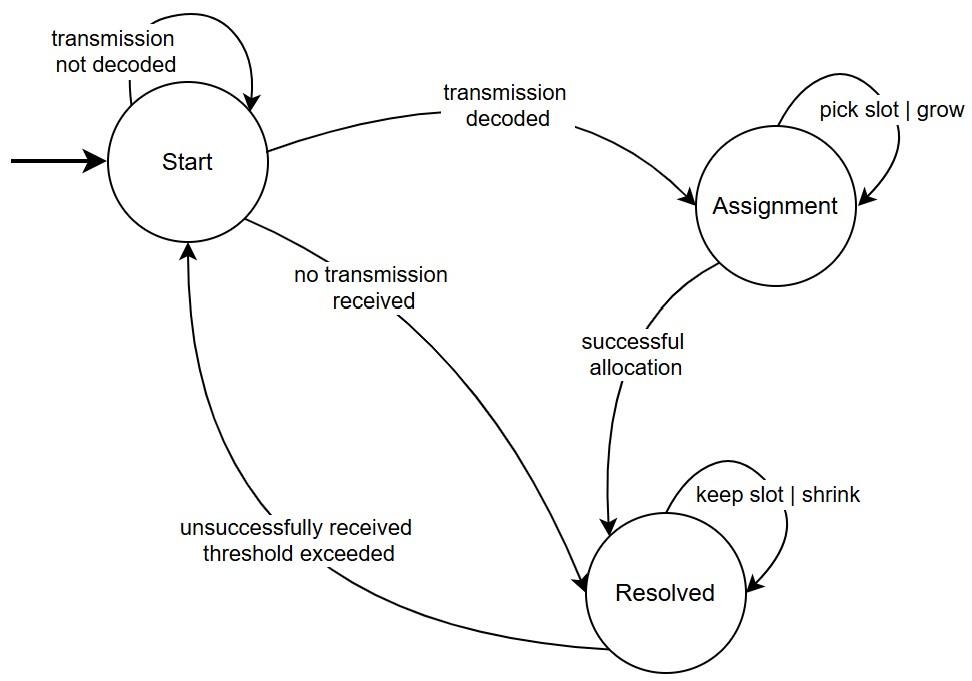}
	\caption{Overview D-STR state diagram}
	\label{fig:dstr_state_diagram}
\end{figure}

Formation \glspl{u} can be in one of three states: the
\textbf{\gls{ss}}, \textbf{\gls{as}}, or \textbf{\gls{rs}}, as seen in
Figure~\ref{fig:dstr_state_diagram}.

A \gls{u} in the \gls{ss} is waiting to receive its first beacon. The first 
UAV is a special case and it will not wait to receive a beacon before 
transmitting its first beacon. Resolving collisions between multiple 
`first' \glspl{u} is out of scope.\footnote{Note that the likelihood of an
    initial beacon collision can be made quite low by using a suitable
    random delay.}

A \gls{u} in the \gls{as} is waiting to self-allocate to a time
slot. A \gls{u} in the \gls{as} will transmit in different time slots
until it successfully picks a slot where its beacon is received by its
neighbors. A \gls{u} in the \gls{as} can trigger a growth in the
superframe size and may react to requests to reduce the superframe
size from \glspl{u} in the \gls{rs}.

A \gls{u} in the \gls{rs} owns a slot and continues
to transmit periodic beacons and is waiting to reduce the superframe
size. A \gls{u} in the \gls{rs} has successfully self-allocated to a
slot and will continue to transmit beacons in that slot. A \gls{u} in
the \gls{rs} may also react to requests to grow the superframe size
from other \glspl{u} that are in the \gls{as}, or it
  can issue a request to shrink the superframe.

\subsection{Self-allocation and Periodic Beacons}
\label{sec:dst_self_allocation}

All formation \glspl{u} begin in the \gls{ss}. A \gls{u}, upon joining
the formation, will listen to the channel for beacons. Assume that
there will always be a single \gls{u} that joins at formation start
up, after which multiple \glspl{u} can join the formation at the same
time. The \gls{dst} protocol does not handle the case where two, or
more, \glspl{u} are simultaneously `first' to join the formation.

If no transmission is received when a \gls{u} joins the formation,
that \gls{u} is the first \gls{u} in the formation. If the \gls{u} is
first, it will adopt the default starting superframe size, claim a
transmission time slot, and enter the \gls{rs}.

If a transmission is received, there are other \glspl{u} in the
formation. In this case, the \gls{u} will wait until it can decode a
beacon before entering the \gls{as}. The \gls{u} will use this beacon
to learn the superframe size and current time slot, and to identify
transmission time slots that are definitely not available using the
following process.

A \gls{u} will consult the record field to determine which time slot
to self-allocate to. The \gls{u} will not self-allocate to any time
slots that have a 1 in the record (indicating successful packet
reception) because this will cause a packet collision. If there are
time slots that have a 0 or 2 in the record, so called
\textbf{available} time slots, the \gls{u} will randomly
select an available time slot to self-allocate to. 
Note that the UAV will only be checking beacon transmissions 
that it receives from neighbors in the preceeding superframe.

The \gls{u} will then transmit a beacon during the time slot it
self-allocates to. If the transmission is successfully received by the
neighborhood, as determined by checking the record of subsequent
beacons, the \gls{u} will claim the time slot and enter the \gls{rs}.
If the transmission is not successfully received by the neighborhood,
the \gls{u} increments its counter, $c$, and remains in the \gls{as},
where it again self-allocates to a randomly selected available time
slot. 

If there are no available time slots, or the \gls{u} has
unsuccessfully self-allocated \gls{ct} times, the \gls{u} will trigger
the addition of time slots to the superframe by transmitting in the
\gls{g} time slot. The counter, c, is set to zero when the \gls{u}
transmits in the \gls{g} time slot or when it enters the \gls{rs}.

Once in the \gls{rs}, a \gls{u} will continue to transmit during its
claimed \gls{ts}. If the \gls{u}'s beacons have not been successfully
received by its neighborhood, the \gls{u} will release the claimed
time slot and return to the \gls{as} with \gls{tsr} probability.

If a \gls{rs} \gls{u} does not receive any beacon record 
indicating the time slot is used, a so called \textbf{silent} time slot, 
it can trigger the potential removal of the silent time slot by transmitting 
in the \gls{s} time slot. Only \gls{rs} \glspl{u} keep track of silent 
time slots and try to shrink the superframe.

\subsection{Growing the Superframe}
\label{sec:dst_growing_superframe}

Two time slots are required for the Growing Mechanism: a \gls{g} slot
and a \gls{gn} slot.  Transmissions in the \gls{g} and \gls{gn} slots
are made with the higher transmit power setting, and therefore will be
received by all \glspl{u} in the formation.  The \gls{g} slot is used
to indicate that the superframe size should increase. There are two
cases where this will happen: the \gls{u} does not observe any
available time slots in the superframe, or the \gls{u} has
unsuccessfully self allocated too many times.  The \gls{gn} slot is
used to achieve global consensus by indicating that multiple \glspl{u}
have transmitted in the \gls{g}. The presence of a transmission in the
\gls{gn} slot indicates that multiple \glspl{u} have transmitted in
the \gls{g} and that \gls{gm} slots should be added to the end of the
superframe. It is not necessary to successfully receive a packet in
the \gls{gn} time slot. Rather, detecting energy in that time slot is
sufficient.

A \gls{u} in the \gls{as} will transmit in the \gls{g} time slot when
it does not observe any available time slots. Recall that time slots
where a \gls{u} has recorded a 0 or 2 are available. This \gls{u}
intends to prompt the addition of 1 time slot to the superframe.

A \gls{u} in the \gls{as} will also transmit in the \gls{g} time slot
when it has unsuccessfully self-allocated \gls{ct} times. This \gls{u}
intends to prompt the addition of \gls{gm} time slots to the
superframe. This increase in the superframe size by \gls{gm} slots
reflects the contention for time slots and will increase the \gls{u}'s
likelihood to successfully self-allocate.

If exactly one \gls{u} transmits in the \gls{g} slot, it is assumed
that the transmission can be decoded by all other formation \glspl{u}
and no \glspl{u} will transmit in the \gls{gn} slot. The superframe
will either grow by 1 time slot or \gls{gm} time slots, depending on
the beacon contents.

However, if multiple \glspl{u} transmit in the \gls{g} slot, the
\glspl{u} that could not decode the beacon (i.e.\ the \glspl{u} that
are aware that more than one \gls{u} is attempting to trigger
superframe growth) will transmit a packet in the \gls{gn} slot.  This
indicates that \gls{gm} slots should be added to the end of the
superframe.  This is done even if the transmissions in the \gls{g}
slot indicate that the superframe should grow by 1 time slot, because
a beacon in the \gls{gn} slot indicates multiple \glspl{u} trying to
trigger a superframe size growth, growing by the \gls{gm} alleviates
the contention over time slots more than simply growing by 1 time
slot.

If a \gls{u} sends or receives a transmission in the \gls{gn} slot, it
will grow the superframe by \gls{gm} slots regardless of whether the
transmission received in the \gls{gn} slot can be decoded or not. As a
result, a formation using \gls{dst} achieves global consensus on the
superframe size.

\subsubsection{Discussion of the growing mechanism}

Consensus is critical for supporting the transmission of successfully
received beacon transmissions.  A lack of consensus affects periodic
beacon transmission, introduces collisions, and makes it difficult for
a \gls{u} to know what superframe size is correct if it receives
conflicting information from different \glspl{u} in its neighborhood.

With the growing mechanism designed for \gls{dst}, negotiation for
growing the superframe occurs within one \gls{sup}'s \gls{g} and
\gls{gn} slots. A \gls{u} formation using the \gls{dst} protocol will
reach a consensus of whether or not to grow the superframe size, and
by what margin, in a maximum of two time slots. This is effective, and
a marked improvement from solutions that rely on the information
propagating through the formation or those that rely on the
information being successfully received by all formation \glspl{u}. It
is of note that our protocol design relies on all formation UAVs
receiving these transmissions, and the transmit power used therefore
influences the geographic spread of a formation that can be supported.
Furthermore, the growing mechanism is designed such that new time
slots are added to the end of the superframe, ensuring that the
existing order of the transmission schedule is unchanged and aiming to
reduce disruption to established periodic transmissions.

\subsection{Shrinking the Superframe}
\label{sec:dst_shrinking_superframe}

Three time slots are required for the Shrinking Mechanism: a \gls{s}
slot, where a \gls{u} proposes to remove a time slot from the
superframe, a \gls{so} slot, where a \gls{u} can object to the removal
of the proposed time slot, and a \gls{sn} slot, which indicates that a
transmission in the \gls{s} time slot was detected but not
successfully received.  Transmissions in the \gls{s}, \gls{so} and
\gls{sn} slots are made with the higher transmit power setting, and
therefore will be received by all \glspl{u} in the formation.

If only one \gls{u} transmits in the \gls{s} slot, the transmission
will be successfully received by all other formation \glspl{u}.
Recall the assumption that the formation is not subject to external
interference.

One case where the superframe may need to shrink is when a \gls{u}
leaves the formation and its \gls{ts} becomes available. Recall from
Table~\ref{tab:dst_beacon_structure} that the beacon contains a
\gls{lf} indicating that the \gls{u} is leaving the formation and that
its time slot is now available. A \gls{u} that is leaving the
formation will transmit its final beacon with a leaving flag. Any
neighbors that are in the \gls{as} will treat this time slot as
available and may choose to transmit during that time slot in the next
superframe.

Another case where the superframe may need to shrink is where a time
slot is not being used by any of the \glspl{u}, a so called 'removable'
time slot. In this case, a \gls{u} in the \gls{rs} may observe that 
it has not received any transmissions in a \gls{ts} for \gls{st} 
superframes. If so, the \gls{u} will presume the \gls{ts} is unused.
Note that a UAV's understanding of the superframe allocation is limited
to its neighborhood. A UAV's record of a locally unallocated time slot 
(i.e.\ an unused slot in the UAV's record) does not mean that the slot
is removable (i.e.\ a globally unallocated time slot).

If a \gls{u} thinks a \gls{ts} is unused, it will prepare to transmit
in the \gls{s} slot. If a transmission is received in the \gls{g} or
\gls{gn} slot, the \gls{u} will not transmit in the \gls{s}
slot and will reset its counter. This is done because time slots
should not be removed while there are still \glspl{u} trying to
self-allocate.

The \gls{so} time slot is used to object to the removal of a proposed
time slot.  Any \gls{rs} \glspl{u} that are able to decode the \gls{s}
beacon will check the proposed time slot. If a \gls{rs} \gls{u} is
self-allocated to the proposed time slot, it will transmit a beacon in
the \gls{so} slot. Any \glspl{u} that are in the \gls{as} will also
transmit a beacon in the \gls{so} slot, irrespective of whether they
can decode the \gls{s} slot beacon. This is done to prevent the
superframe shrinking while there are \glspl{u} still looking for an
available time slot.

If there is no signal energy in \gls{so} slot, the proposed \gls{ts}
will be removed from the superframe. All time slots after the removed
\gls{ts} will shift by one time slot. For example, if \gls{ts} 3 is
removed and a \gls{u} had previously self-allocated to \gls{ts} 5, the
\gls{u} will now self-allocate to \gls{ts} 4. If one or more \glspl{u}
transmit in the \gls{so} slot, the proposed time slot will not be
removed and the superframe will remain as is.

If multiple \glspl{u} transmit in the \gls{s} slot, all \glspl{u} that
cannot decode the transmission will transmit in the \gls{sn} slot to
indicate that the \gls{s} slot transmission was not received
successfully.

If there are transmissions in the \gls{s} and \gls{sn} slots only, any
\gls{u} that transmitted in the \gls{s} slot will reattempt its
transmission using a truncated exponential backoff process. Each
\gls{u} maintains a count, $s$, of unsuccessful transmissions in the
\gls{s} slot. The backoff period, in superframes, is selected from a window of size
$[0,
2^{min\{s,S\}}-1]$.

If there are transmissions in the \gls{s} and \gls{so} slots, any
\gls{u} that transmitted in the \gls{s} slot will not reattempt its
transmission irrespective of any transmissions in the \gls{sn}
slot. Instead, it will reset its \gls{st} counter to zero, and will
add the proposed \gls{ts} to its \gls{fs}.

\subsubsection{The \gls{fs}}

Every \gls{u} has a \gls{fs}. This keeps track of any \gls{ts} the
\gls{u} has unsuccessfully attempted to remove. For instance, a
\gls{u} proposes to remove transmission \gls{ts} 5 and receives an
objection. \Gls{ts} 5 is added to the \gls{u}'s \gls{fs}. If a
\gls{ts} is in its \gls{fs}, a \gls{u} will not propose its removal.

The \gls{fs} has a timeout, \gls{to}, that determines how frequently a
\gls{u} can propose the removal of a given \gls{ts}. Each entry in the
\gls{fs} has a timer, $t_{n}$, that counts the superframes that an
entry has been in the \gls{fs}. When a time slot entry, $n$, is added
to the \gls{fs}, its time $t_{n}$ is initialized to zero. Every
superframe, $t_{n}$ is incremented. Entries that have been in the
\gls{fs} for \gls{to} superframes time out and will be removed from
the \gls{fs}. Because negotiating the removal of a time slot can occur 
each superframe, these values are configured relative to the superframe 
size. 

The \gls{fs} is updated whenever a \gls{ts} is removed from the
superframe. For example, if time slot 5 is removed from the superframe
and the \gls{fs} has entries for time slots 3, 5, and 7, the entry for
the removed time slot is erased and any higher time slot indices are
updated. In this case, the \gls{fs} entries are now 3 and 6.

\subsubsection{Discussion of the shrinking mechanism}

It is possible for all formation \glspl{u} to transmit in the \gls{s}
time slot and therefore potentially lose consensus over the
superframe. To solve the problem of all \glspl{u} transmitting in the
\gls{s} time slot, the protocol will `map' a \gls{u}'s initial backoff
to its order in the superframe. Say that \glspl{u} $1, 2, 3, 5, 6$ all
wish to transmit in the \gls{s}. \gls{u} $1$ will have a 1 superframe
backoff, \gls{u} $2$ will have a 2 superframe backoff, \gls{u} $3$
will have a 3 superframe backoff, \gls{u} $5$ will have a 4 superframe
backoff, and \gls{u} $6$ will have a 5 superframe backoff. This fixes
the problem by ensuring that not all formation \glspl{u} choose the
same backoff period. The backoff is relative to the superframe size, 
and therefore the shrink process can become longer as the superframe 
size grows.

With the proposed shrinking mechanism, negotiation for removing a time
slot occurs within a minimum of one three time slots.
However, this process may take longer if the \glspl{u}' transmissions
in the \gls{s} slot collide and no objection is given in the \gls{so}
slot. In the worst-case this takes $O(n)$ superframes, where $n$ is
the size of the neighborhood.

The design of the \gls{dst} protocol is such that the protocol is
faster to grow than to shrink. This is because the aim is firstly to
achieve a schedule that supports periodic updates, so that formation
\glspl{u} do not experience physical collisions, and then to reduce
the update period by eliminating any unused time slots. With this aim,
\gls{dst} expressly prioritizes keeping a time slot over removing it
if there is a possibility that it is being used for periodic
updates. Further, the design of the shrinking mechanism maintains the
integrity of periodic updates. Even when a time slot is removed, the
order of transmissions is unchanged. In the absence of other changes
in the superframe, such as new \glspl{u} joining the formation, this
guarantees the continuation of periodic beacons that are successfully
received.

\subsection{Detecting Reception Success}
\label{sec:dst_collision_detection}

Detecting if a transmission was successfully received is important
when a \gls{u} is determining whether to continue transmitting in a
given \gls{ts}.

To evaluate the success of its self-allocation, an \gls{as} \gls{u}
needs to determine if its transmission was successfully received. If
the transmission was successfully received by all neighbors, the
self-allocation was successful and the \gls{u} can claim the time slot
and enter the \gls{rs}. If the transmission was not successfully
received (i.e.\ if even a single neighbor beacon indicated that the
transmission was not successfully received), the self-allocation was
also unsuccessful and the \gls{u} will remain in the \gls{as} and
continue to self-allocate.

Similarly, a \gls{rs} \gls{u} needs to determine if its transmission
is successfully received to evaluate its time slot claim. Each
\gls{rs} \gls{u} keeps a counter $F$ to keep track of failed beacon
receptions. The \gls{u} does not tolerate multiple unsuccessfully
received transmissions. If the transmission is successfully received
by all neighbors, the \gls{u} will set $F=0$ and continue to transmit
in its claimed time slot. If the transmission is not successfully
received by all neighbors more than once, the \gls{u} will increment
$F$. If $F>0$ the \gls{u} will release its claimed time slot and
return to the \gls{as} with a \gls{tsr} probability $p$.

As discussed previously, each \gls{u} maintains a record of received
transmissions and the bits in this record are mapped to time
slots. Recall that, if a \gls{u} receives a transmission in a time
slot and can decode it, the corresponding bits in its record will be
set to `01'.

Upon receiving a beacon, the \gls{u} will use the beacon's record
field to determine if its last transmission was successfully
received. If the bits corresponding with the \gls{u}'s self-allocated
slot are set to `01', the transmission was successfully received.  If
all its neighbors report successfully receiving its last transmission,
an \gls{as} \gls{u} will claim the time slot and enter the \gls{rs}. A
\gls{rs} \gls{u} will reset its counter $F$ to zero. Otherwise, an
\gls{as} \gls{u} will self-allocate to an available time slot, and a
\gls{rs} \gls{u} will increment its counter $F$. Beacon record fields
update when the superframe size changes, and beacons sent subsequent
to a change in superframe size will contain a larger or smaller record 
field to reflect this. 

It is possible for a \gls{u} to have not yet heard from all its
neighbors. Note that a \gls{u} does not know how many neighbors it has
up front and may not receive beacons from all its neighbors in a given
superframe. Therefore, a \gls{u} might think that its transmission was
successfully received, and claim a time slot, even when a neighbor
does not successfully receive the transmission. In the following
superframe(s), however, the neighbor will transmit and the \gls{u}
will be informed that its transmission was unsuccessfully
received. Additionally, the requirement adopted in this work, that
beacons must be successfully received by all neighbors, is rather
strict and may not be suitable for very dense neighborhoods and very
large superframe sizes.

\section{Results}
\label{sec:results}

The \gls{dst} protocol was implemented in Python. Simulations were run
$10,000$ times. Protocol behavior was validated during simulation,
with errors being raised if unexpected situations arise (e.g.\ if a
\gls{u} believes it is receiving transmissions from an invalid
transmitter, or all \glspl{u} transmit in the same time slot, or a
\gls{u} believes it can transmit and receive at the same time). No
errors were raised in any of the simulations. Unit tests were also
created to test the behavior of the protocol (including testing that
\glspl{u} resolve after growing if each \gls{u} selects a unique time
slot and testing that non-establishing \glspl{u} wait to receive a
beacon before transmitting)

The allocation is also validated, to ensure that the formation
\glspl{u} have an accurate understanding of the state of their
transmissions. This is done by checking, for each time slot, that the
SINR for each neighbor \gls{u} is acceptable. Simulations can either
run until resolution is achieved (i.e.\ all \glspl{u} are in the
\gls{rs}) or convergence is achieved (i.e.\ all \glspl{u} are in the
\gls{rs} and there are no removable time slots). 
The final allocation (upon resolution and/or convergence) is also validated, 
and there were no instances where the final allocation was invalid. 
In addition, there were no instances where a converged formation had 
a superframe with any unallocated time slots.

\subsection{Sensitivity Study}
\label{sec:dstr_results_sensitivity}

A sensitivity study was conducted to get some insight into the effects
that the parameters (independent variables) have on particular
outcomes (dependent variables) with respect to the effectiveness of
the protocol. The fixed parameter settings chosen for
this study are outlined in Table~\ref{tab:spatial_reuse_parameters}.

\begin{table}[]
\centering
\begin{tabular}{lll}
\hline
\multicolumn{3}{c}{\textbf{Protocol and Channel Parameters}}                                                       \\ \hline
\multicolumn{1}{l}{Parameter} & \multicolumn{1}{l}{Value} & \multicolumn{1}{l}{Comment}                            \\ \hline
$\lambda$   & 0.125      & wavelength                           \\ 
$N$         & -101 dBm   & total noise power                    \\ 
$\gamma$    & 2          & path loss exponent                   \\ 
$PL_0$      & 40 dB      & path loss at reference distance      \\
$\Gamma$    & 15 dB      & required SINR                        \\ 
            & 512 B      & packet size                          \\ \hline \hline
\\ \hline
\multicolumn{3}{c}{\textbf{Formation Deployment Parameters}}    \\ \hline
\multicolumn{1}{l}{Parameter} & \multicolumn{1}{l}{Value} & \multicolumn{1}{l}{Comment}   \\ \hline
$r$         & 10 m       & distance between formation UAVs      \\
$\Delta$    & 10m        & safety radius                        \\ \hline \hline
\end{tabular}
  \caption{The parameters used, based on transmission over $2.4$ GHz WiFi. Friis free space path loss model is used.}
  \label{tab:spatial_reuse_parameters}
\end{table}

\begin{figure}
    \centering
    \includegraphics[width=0.75\linewidth]{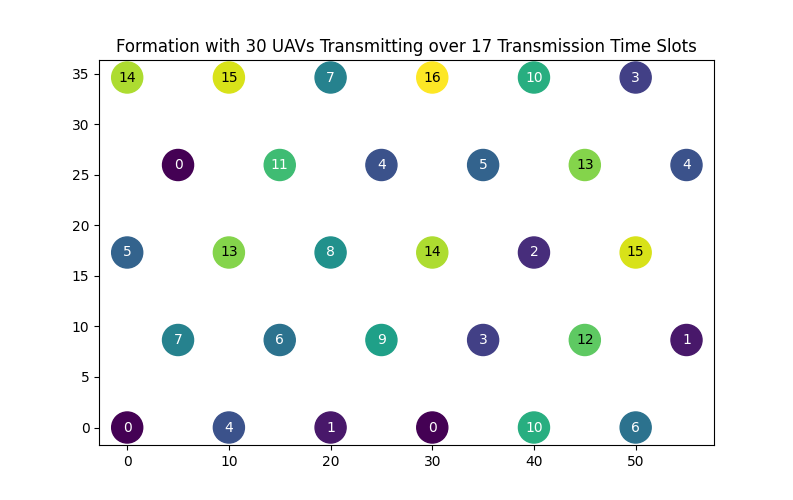}
    \caption{An example of the formation structure adopted for the sensitivity study. The circles represent the positions of the \glspl{u}. The color and number assigned to a circle indicate the transmission time slot used by the \gls{u} in that position.}
    \label{fig:dst_sensitivity_study_formation_structure}
\end{figure}

The \glspl{u} are within a rectangular formation and follow a hexagonal structure, with a 10m separation between \glspl{u}, as shown in Figure~\ref{fig:dst_sensitivity_study_formation_structure}. The binary settings (lower and upper) chosen for individual parameters are outlined in Table~\ref{tab:dstr_sensitivity_study_parameter_settings}.

\begin{table}
    \centering
    \begin{tabular}{l|c|c}
        Parameter                           & Lower Setting     &  Upper Setting    \\ \hline
        Number of UAVs (U)                     & 10                & 100               \\
        Time Slot Retention Probability (TSR)     & 0.75              & 0.95              \\ 
        Collision Threshold (CT)                & 3                 & 7                 \\ 
        Growth Margin   (GM)                    & 3                 & 9                 \\ 
        Shrink Threshold  (ST)                  & 5                 & 10                \\
    \end{tabular}
    \caption{Settings for testing the effect of the parameters on the behavior of the \gls{dst} protocol.}
    \label{tab:dstr_sensitivity_study_parameter_settings}
\end{table}

An ordinary least squares (OLS) linear regression model was used. OLS minimizes the sum of squared residuals (i.e.\ the distance between the data point and the regression line) to find the best fit~\cite{james2023introduction}. The analysis was conducted using the statsmodels package for Python~\cite{seabold2010statsmodels}. 
The reported coefficient values, if significant, can be used to determine the effect that a parameter setting has on a dependent variable.
If the reported $P > |t|$ is smaller than $0.05$, the associated coefficient value is significant and can explain a change in the dependent variable.

The outcomes of interest are:
\begin{itemize}
\item Time to convergence (normalized to reflect the number of `rounds'). 
    This refers to the time until the final superframe state is reached, 
    where all UAVs are resolved and there are no removable time slots.
    Rounds are calculated as the number of time slots until convergence
    divided by the formation size.
  \item Final superframe size (as the superframe length in time slots).
  \item Resolution time (normalized to reflect the number of `rounds'). 
    This refers to the time until the final allocation order is
    reached, where all UAVs are resolved but there
    can be removable time slots.
  \item The number of control packets (normalized to reflect the
    number of control frames generated by a UAV per round). This is 
    calculated as the total number of control packets per \gls{u} in a given round.
  \item UAVs per time slot (once convergence is reached). This value 
    indicates the level of spatial reuse achieved.
  \item Removed slots (total). This value reflects the number of extra time slots 
  that are removed by the shrinking mechanism between resolution and convergence.
\end{itemize}

Overall, the outcomes can be explained by the parameter settings. The
R-squared values are high, as seen in Table~\ref{tab:dst_rsquared},
which indicates that the dependent variable values can be explained by
changes in the parameter settings. For almost all outcomes, the R-squared
values are greater than $0.9$. The only exception is for the number of 
removed slots, which has a $\sim 0.7$ R-squared value. The Adjusted R-squared 
value and R-squared value are equal, indicating that all the parameters chosen 
are contributing to the dependent variable values~\cite{Assuncao1993Introduction}. 
While an R-squared value always increases when the number of independent variables
increases, which can result in artificial inflation of the R-squared value, 
the Adjusted R-squared value penalizes the inclusion of irrelevant independent 
variables that do not contribute to explaining the dependent variable.

\begin{table}[]
    \centering
    \begin{tabular}{l|c|c}
    Outcome                     &  R-squared & Adjusted R-Squared \\ \hline
    Convergence Time            & 0.945 & 0.945 \\
    Resolution Time             & 0.936 & 0.936 \\
    Final Superframe Length     & 0.964 & 0.964 \\
    Number of Control Packets   & 0.917 & 0.917 \\
    UAVs per Time Slot          & 0.965 & 0.965 \\
    Removed Slots               & 0.713 & 0.713 \\
    \end{tabular}
    \caption{The R-squared and adjusted R-squared values of the OLS regression for the tested outcomes.}
    \label{tab:dst_rsquared}
\end{table}

In general, increasing or decreasing a given parameter setting has the
same effect on all the tested outcomes (albeit by differing
margins). For example, increasing the TSR setting results in
unfavorable outcomes. It decreases the number of \glspl{u} per time
slot, and increases the final superframe size, time to convergence,
and resolution time. The TSR also always has the largest coefficient
value. Therefore, decreasing the TSR will result in favorable outcomes
overall.

Similarly, increasing the U setting increases all the
outcomes. However, the associated coefficient values are all
relatively small. For example, each additional \gls{u} adds a
negligible 0.005 control packets per superframe overhead. The largest
increase is seen in the final superframe size, which only increases by
0.021 time slots for each additional \gls{u}.

The final superframe size is of interest because it affects the update delay, 
and thereby the level of certainty a \gls{u} has about the position of its neighbors, 
once the formation has reached convergence. Parameters that have a significant effect 
(a P value $< 0.05$) are reported in Table~\ref{tab:dstr_final_superframe_size_coefficients}. 
All parameters and interactions in the table have a P value $< 0.01$, 
with the exception of the interaction between TSR and CT, which has  a P value of 0.015.

\begin{table}[]
    \centering
    \begin{tabular}{l|c}
        Parameter & Coefficient \\ \hline
        U  &       0.2069  \\
        TSR &       6.5579 \\
        CT &       0.7864  \\
        GM &       0.5329  \\
        U * TSR &      -0.0517  \\
        U * CT &      -0.0076   \\
        U * GM &       0.0149   \\
        TSR * CT &      -0.5975  \\
        TSR * GM &      -0.5442   \\
        CT * GM &      -0.0354   \\
    \end{tabular}
    \caption{OLS regression results for the final superframe size. Coefficients with significant P values (>0.05) are reported.}
    \label{tab:dstr_final_superframe_size_coefficients}
\end{table}

The time to resolution is of interest because it indicates the time taken
until periodic transmissions from all formation UAVs are successfully received. 
At this stage, there may still be removable time slots and the update delay
can possibly be reduced. The coefficients of the parameters and interactions
with significant effect are captured in Table~\ref{tab:dst_resolution_time_coefficients}.

\begin{table}[]
  \centering
  \begin{tabular}{l|c}
      Parameter & Coefficient \\ \hline
      U  &       0.1246 \\
      TSR &     1.5234 \\
      CT &       0.2886 \\
      GM &      0.4999   \\
      U * TSR &     -0.0258    \\
      U * CT &     -0.0098    \\
      U * GM &     0.0109    \\
      TSR * CT &    -0.1098    \\
      TSR * GM &    -0.0900      \\
      CT * GM &      -0.0403    \\
  \end{tabular}
  \caption{OLS regression results for time to resolution (normalized). Coefficients with significant P values ($>0.05$) are reported.}
  \label{tab:dst_resolution_time_coefficients}
\end{table}

The time to convergence is of interest because it indicates the time taken 
until periodic transmissions from all formation UAVs are successfully received 
and the update delay is reduced. A smaller time to convergence means that 
a formation more quickly establishes the final transmission schedule. 
The coefficients of the parameters and interactions with a significant effect 
are captured in Table~\ref{tab:dst_convergence_time_coefficients}. The coefficients 
for U and its interactions all have a standard error less than or equal to 0.004. 
The TSR parameter has a standard error of 0.815. The CT and GM parameters 
(and their interactions) all have a standard error less than or equal to 0.087. 

\begin{table}[]
    \centering
    \begin{tabular}{l|c}
        Parameter & Coefficient \\ \hline
        U  &       0.1826  \\
        TSR &       3.6316 \\
        CT &       -0.1745  \\
        GM &       0.1578  \\
        U * TSR &      -0.0234  \\
        U * CT &      -0.0130   \\
        U * GM &       0.0269   \\
        U * ST &       -0.0007   \\
        TSR * CT &      -0.1971  \\
        TSR * GM &      -0.4180   \\
        CT * GM &      -0.0404   \\
    \end{tabular}
    \caption{OLS regression results for time to convergence (normalized). Coefficients with significant P values ($>0.05$) are reported.}
    \label{tab:dst_convergence_time_coefficients}
\end{table}

\subsubsection{Discussion}

Increasing the CT by one unit reduces the time to convergence by 0.17
rounds and increases the final superframe size by 0.79 time slots on
average. This indicates that increasing the number of attempts a UAV
makes to find a time slot where its transmissions are successfully
received, increases the likelihood of finding a successful final
allocation in fewer rounds, at the expense of superframe size.

Lowering the TSR reduces the final superframe size by 6.56 time slots
and the time to convergence by 3.63 rounds on average. 
This indicates that making the allocation more adaptable increases the
likelihood of finding a successful final allocation in fewer rounds
with less need to grow the superframe size.

The time to resolution offers an indication of how well contention is
managed, because a smaller time to resolution indicates fewer rounds
until all \glspl{u} are in the \gls{rs}. A one unit increase in the GM
increases the final superframe size by 0.53 time slots and the
resolution time by 0.5 rounds, indicating that every two unit GM
increase likely makes the superframe one time slot longer. Therefore,
a smaller GM setting should be chosen if the main aim is to reduce the
superframe size.

The shrink parameters do not have a significant effect on the time to
convergence, with the ST and FST parameters having p values greater
than 0.05. This indicates that, for the formation sizes tested in this
study, the number of rounds until convergence is not limited by the
shrinking mechanism and the parameter settings it uses.

The formations tested had between 0-37 removable time slots
(i.e.\ transmission time slots in which no \glspl{u} transmits
beacons) when resolved. The protocol was able to remove all 
extra time slots that were present when the formation resolved 
and the formations always reached convergence.

The overhead for a formation using the \gls{dst} protocol is
essentially stable. The formation size has a small effect on the
overhead, with each additional formation \gls{u} adding an overhead of
0.0049 packets per \gls{u}. This indicates that increasing the
formation size does not generate the need for much additional overhead
to reach a final allocation. The total normalized overhead ranged from
$\sim0.02-0.8$ for all combinations of parameter settings.

A one unit increase of the GM and ST parameter settings reduces the
number of control packets by 0.03 and 0.01 packets per \gls{u}
respectively. The p-value for the GM and ST parameters is greater than
the threshold adopted ($0.05$) and therefore these results are
statistically significant. This result indicates that the overhead can
be very slightly improved by smaller $GM$ and $ST$ settings. However,
while the reported results are statistically significant, a difference
of 0.03 or 0.01 is too small to have a practical effect.

The time slot reuse factor is highest (conversely, the superframe size
is smallest) when the CT is at the higher setting and the GM is at the
lower setting. This is to be expected, as a higher CT will force
\glspl{u} to repeatedly try different transmission time slots before
reducing contention. The lower GM setting reduces the available time
slots to choose from. This coupled with the higher CT means that more
attempts are made to find a feasible allocation with a smaller
superframe. When U = 10, there are on average 1.10 \glspl{u} per time
slot (min = 1.07, max = 1.13). The lower reuse factor is due to the
small geographic spread of the formation. When U = 100, there are on
average 3.79 \glspl{u} per time slot (min = 3.07, max = 4.69). The
reuse factor is higher due to the increased geographic spread of the
formation. The reuse factor increases as the geographic spread of the
formation increases and very large formation sizes experience the
greatest benefit from spatial reuse.

\subsubsection{The Default Superframe Size}
\label{sec:dstr_dss_results}

\begin{figure}
    \centering
    \includegraphics[width=0.8\linewidth]{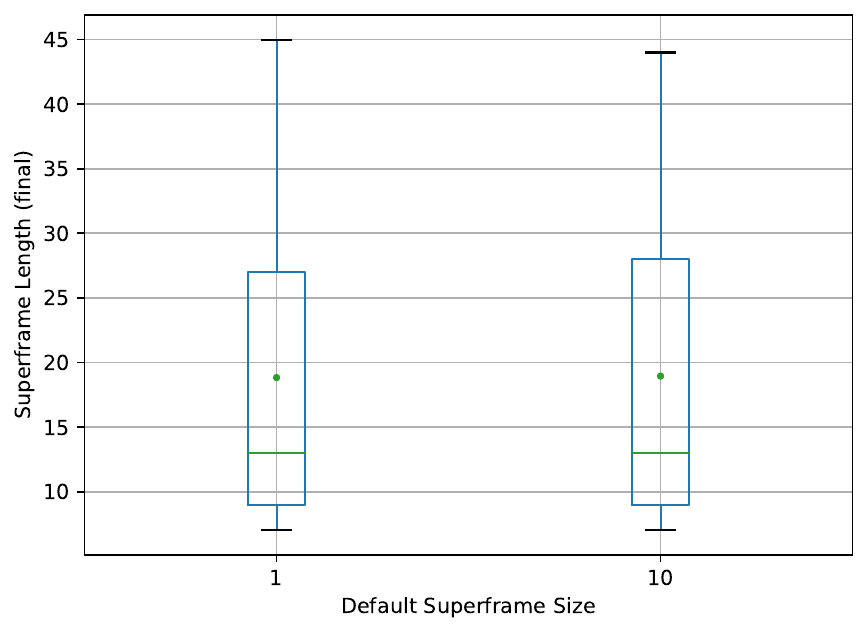}
    \caption{The final superframe length based on the default superframe size.}
    \label{fig:superframe_dss_box}
\end{figure}

\begin{figure}
    \centering
    \includegraphics[width=0.8\linewidth]{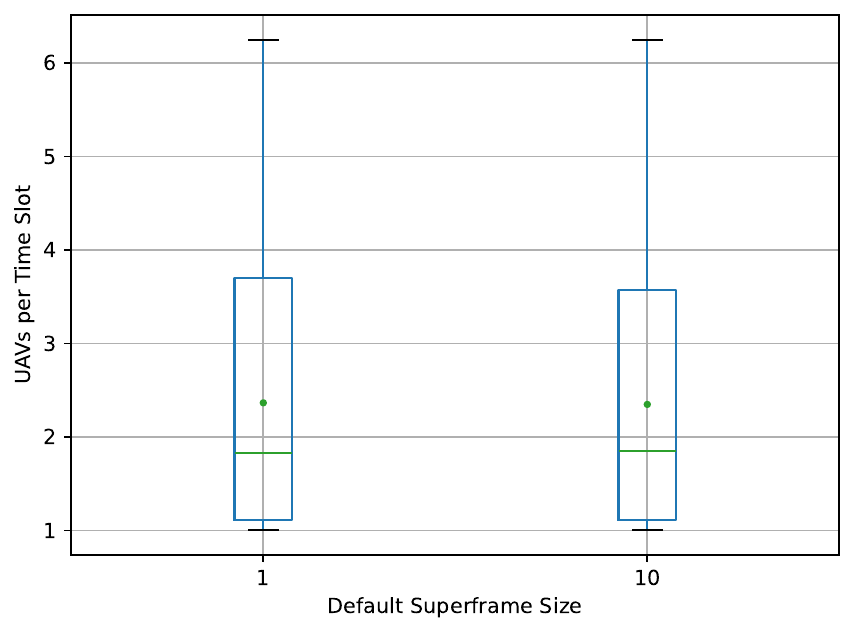}
    \caption{The level of spatial reuse, captured as the number of \glspl{u} per time slot, based on the default superframe size.}
    \label{fig:reuse_dss_box}
\end{figure}

The \gls{dss} was tested with 1 and 10 as the lower and higher settings respectively. All other settings are as defined in Table~\ref{tab:dstr_sensitivity_study_parameter_settings}. The default superframe size does not appear to have much effect on the final superframe length, or by extension the level of spatial reuse, as seen in Figures~\ref{fig:superframe_dss_box} and~\ref{fig:reuse_dss_box}.

\begin{figure}
    \centering
    \includegraphics[width=0.8\linewidth]{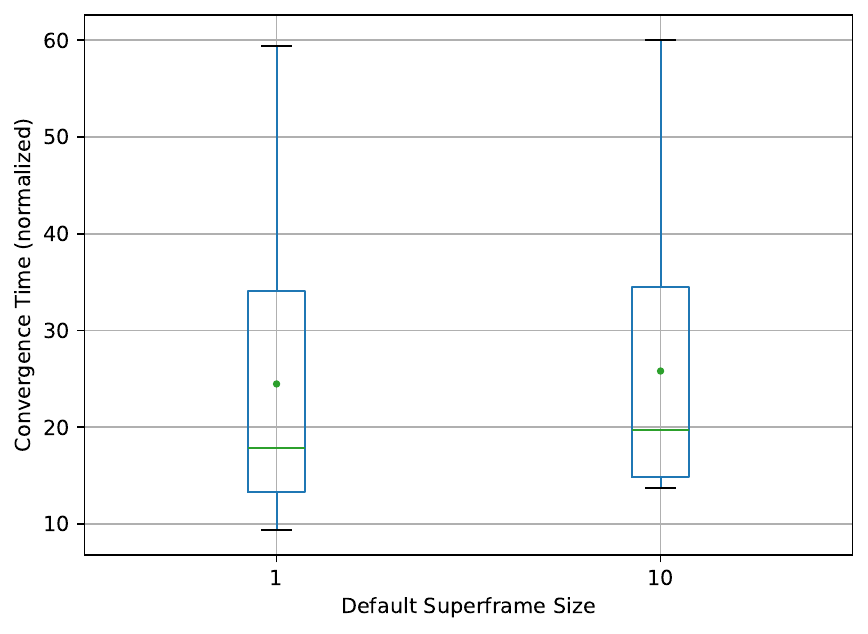}
    \caption{The normalized convergence time at the upper and lower default superframe size.}
    \label{fig:dst_dss_convergence}
\end{figure}

As seen in Figure~\ref{fig:dst_dss_convergence}, the lower \gls{dss}
setting results in shorter normalized time to convergence.  While this
result is statistically significant ($p\leq 0.05$), the impact on time
to convergence is slight. Maximum convergence times are nearly equal
and the average and median time to convergence only increases by
$\sim$ 1-2 rounds. The higher \gls{dss} setting reduces the number of
superframes until resolution and convergence.

\subsubsection{Shrinking Mechanism Parameters}
\label{sec:dst_shrink_mechanism_parameters}

The ST and FST parameters used for the shrinking mechanism were also
tested to determine their effect. The \gls{st}, which a \gls{u} uses
to evaluate silent \glspl{ts}, was set to 5 and 10 superframes, which
determines how long a \gls{u} waits before attempting to remove an
`extra' time slot. The FST, which determines how long an entry remains
in a \gls{u}'s \gls{fs}, was set to 10 and 30 superframes. The other
parameter settings were as outlined in
Table~\ref{tab:dstr_sensitivity_study_parameter_settings} and
Section~\ref{sec:dstr_dss_results}.

The ST and FST settings did not have a statistically significant effect on the time to convergence, final superframe size, resolution, or the number of \glspl{u} per time slot. Furthermore, the FST setting did not have a statistically significant effect on the overhead.

This outcome was expected. The shrinking mechanism occurs after resolution, when the superframe size and number of \glspl{u} per time slot have already been established. Hence the shrinking mechanism parameter settings do not have much influence over these outcomes.

\begin{figure}
    \centering
    \includegraphics[width=0.8\linewidth]{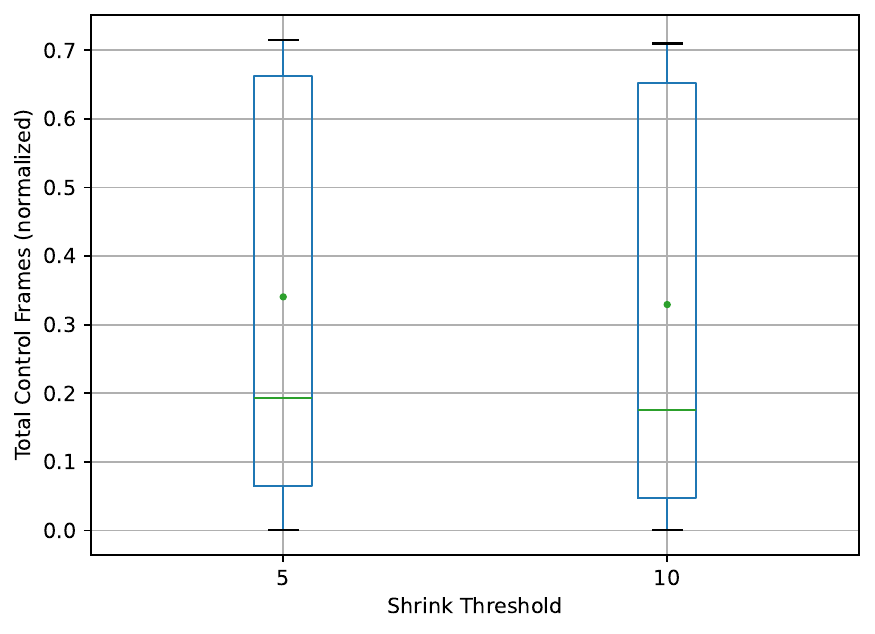}
    \caption{The overhead based on the shrink threshold setting.}
    \label{fig:overhead_st_box}
\end{figure}

The ST parameter had a small but significant effect on the number of control packets (normalized) that were generated. The ST had a coefficient of -0.0036 and as seen in Figure~\ref{fig:overhead_st_box} the higher ST value reduced the mode very slightly. The distributions are quite similar, with both having a left skew and essentially equivalent ranges and mean values. 

\subsubsection{Validating the Shrinking Mechanism}

\begin{figure}
    \centering
    \includegraphics[width=0.7\columnwidth]{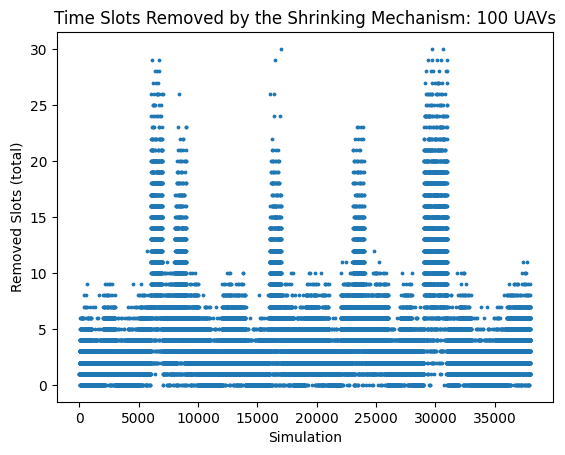}
    \caption{The number of time slots removed by the shrinking mechanism across all simulations with a formation size U=100.}
    \label{fig:dstr_removed_slots_100}
\end{figure}

\begin{figure}
    \centering
    \includegraphics[width=0.7\columnwidth]{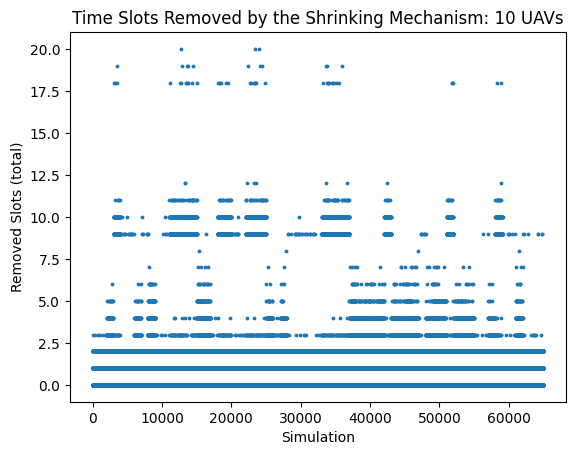}
    \caption{The number of time slots removed by the shrinking mechanism across all simulations with a formation size U=100.}
    \label{fig:dstr_removed_slots_10}
\end{figure}

As seen in Figure~\ref{fig:dstr_removed_slots_100}, the shrinking 
mechanism removed anywhere from 0 to 37 time slots, depending on 
how many extra time slots were present in the superframe upon 
resolution. This demonstrates that the shrinking mechanism does 
remove extra time slots between resolution and convergence.

In the simulations, unexpected outcomes, such as the erroneous removal
of time slots or the superframe length shrinking before resolution,
would raise exceptions. This never occurred, indicating that the
protocol does behave as intended. The number of removed slots shown in
Figures~\ref{fig:dstr_removed_slots_10}
and~\ref{fig:dstr_removed_slots_100} was never negative. This
indicated that the superframe size at convergence was always of
shorter or equal duration to the superframe size at resolution, and
further confirmed that the protocol behaved as intended.

With the right combination of parameter settings, it was possible to
reduce the number of silent \glspl{ts} in a given allocation, thereby
reducing the update delay for a formation upon resolution.  This is
advantageous because the shrinking mechanism is designed to prevent
the erroneous removal of a time slot that is `in use'. Therefore, all
\glspl{u} must achieve consensus on the slot to be removed, which can
potentially be lengthy if transmissions collide and multiple backoffs 
are required.

However, some parameter settings increase the number of silent time
slots while providing other important benefits:
\begin{itemize}
\item Increasing the GM decreases the number of \glspl{u} per time
  slot and increases the final superframe size, time to convergence,
  and resolution time. However, it also has the benefit of decreasing
  the number of control packets.
\item Increasing the CT decreases the number of \glspl{u} per time
  slot, and increases the final superframe size and resolution
  time. However, increasing the CT also has the benefit of decreasing
  the time to convergence and the number of control packets.  When a
  higher CT setting is used, the superframe will grow less often and
  there will be fewer `in-use' and `extra' time slots.  Removing
  `extra' time slots from the superframe is the most costly operation
  because of the protocol's aim to maintain successful transmissions
  after resolution by preventing time slots that are in use from being
  removed from the superframe. Therefore, the time to convergence is
  shorter when there are fewer `extra' time slots in the superframe at
  resolution.
\end{itemize}


\subsection{Comparing D-STR with Centralized Allocation}
\label{sec:dst_comparison_cent}

The \gls{dst} protocol was used on large formation sizes to compare
the final superframe size with the allocation determined by our
centralized spatial reuse scheme~\cite{samandari2022tdma}.

Our centralized spatial reuse scheme was designed to reduce the
superframe size, and by extension the update delay, by removing
the need for each UAV to be allocated to a unique time slot. The
allocations provided by our scheme are feasible and satisfy SINR
requirements that indicate successful reception of periodic beacons.

\begin{figure}
  \centering
  \includegraphics[width=0.6\textwidth]{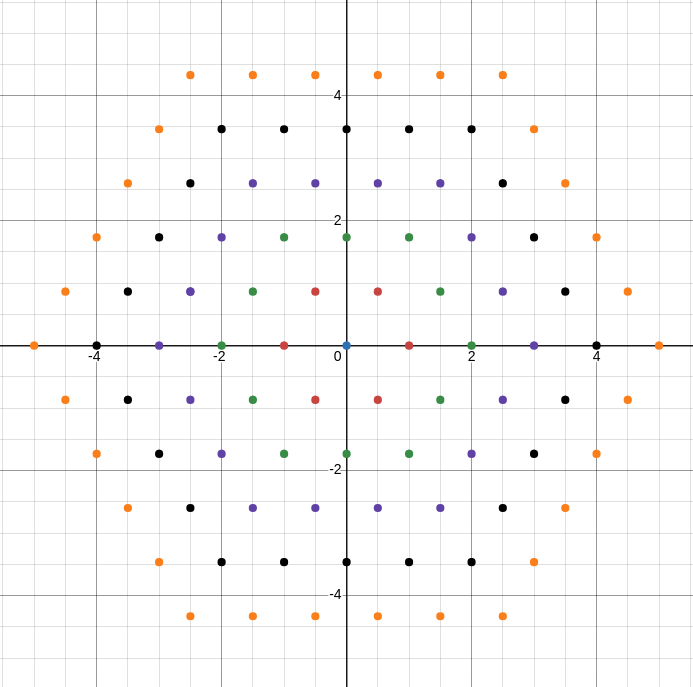}
  \caption{An example of the formation structure used to compare with the centralized scheme.}
  \label{fig:5tier_formation}
\end{figure}

The D-STR protocol was run on formations where the \glspl{u} are in 
hexagonal `rings' around a center \gls{u} with a separation of 10 m (as seen in
Figure~\ref{fig:5tier_formation}). Simulations were run 10 times. The
settings used for all simulations were $TSR = 0.05$, $CT = 7$,
$ST = 3$. The final superframe size results are captured in
Table~\ref{tab:dst_large_formations}. The reuse factor (i.e.\ number of
UAVs per time slot) and the final superframe length for both the
centralized scheme and \gls{dst} are captured in
Table~\ref{tab:dst_central_compare}.

The superframe length achieved by the centralized scheme is
consistently smaller than the \gls{dst} protocol, but the \gls{dst}
protocol can achieve results within a modest factor. For example, the
centralized scheme demonstrated that formations of size 4 - 8 rings
can be supported by a superframe with 91 time slots. The average
superframe length with \gls{dst} is only 16.7 time slots longer than
the centralized scheme for a formation with 4 rings. However, the
average superframe length with \gls{dst} is 63.9 time slots longer for
a formation with 8 rings. The final superframe length and number of
UAVs per time slot (i.e.\ reuse factor) are comparable for the
smallest formation size that is supported by a given allocation tile
size. There is a $\sim3.56$ times increase in the number of UAVs from
a 4 ring to an 8 ring formation. With \gls{dst}, the superframe size
only increases $\sim1.44$ times from a formation with 4 to 8 rings.

\begin{table}[]
    \centering
    \begin{tabular}{|c|c|c|c|c|c|c|c|c|c|c|}
    \hline
    U       & MEAN  & STD   & VAR   & MIN   & MAX   & 25TH     & 50TH       & 75TH      \\ \hline
    271	    & 49.1	& 1.64	& 2.69	& 47	& 52	& 47.25	   & 50	        & 50	    \\ \hline
    1261	& 76.3	& 2.15	& 4.61	& 73	& 81	&75	       & 76.5	    & 77	    \\ \hline
    2269	& 85.7	& 3.44	& 11.81	& 79	& 91	& 85	   & 85.5	    & 86.75	    \\ \hline
    5677	& 107.7	& 3.63	& 13.21	& 102	& 114	& 105	   & 107.5	    & 110.5	    \\ \hline
    19927	& 154.9	& 6.09	& 37.09	& 147	& 169	& 152.25   & 153.5	    & 157.75	\\ \hline
    \end{tabular}
    \caption{Final superframe size generated by \gls{dst} protocol.}
    \label{tab:dst_large_formations}
\end{table}

\begin{table}[]
\centering
\begin{tabular}{|c|c|c|c|c|c|c|c|c|}
\cline{2-5}
\multicolumn{1}{c}{}& \multicolumn{2}{|c|}{UAVs/Time Slot} & \multicolumn{2}{|c|}{Superframe Length} \\ \cline{1-5} 
rings           & Centralized   & \gls{dst}     & Centralized   & \gls{dst}   \\ \hline
 1              & 7             & 5.5           & 37            & 49.1      \\ \hline
 2              & 19            & 16.5          & 61            & 76.3     \\ \hline
 3              & 37            & 26.5          & 61            & 85.7     \\ \hline
 4              & 61            & 52.7          & 91            & 107.7     \\ \hline
 8              & 217           & 128.6         & 91            & 154.9    \\ \hline
\end{tabular}
\caption{Results comparing the proposed centralized scheme and the \gls{dst} protocol.}\label{tab:dst_central_compare}
\end{table}

\subsection{Comparing D-STR with Single-Hop Allocation}
\label{sec:dst_comparison_single_hop}
The \gls{dst} protocol was used on the single-hop case. This was done
to compare the convergence time with our \gls{dar} protocol and to
confirm that the \gls{dst} protocol achieves competitive performance
in the single-hop case despite its higher complexity.

D-ART is a novel, distributed, TDMA-based protocol that is designed
to support the single-hop exchange of safety messages across a 
fully-connected UAV formation~\cite{samandari2023distributed}. 
D-ART aims to find the best possible allocation schedule 
(the smallest possible superframe size) given a population of 
unknown size. As with our D-STR protocol, this requires eliminating 
removable time slots from the superframe. However, because the 
use case is single-hop communication in a fully-connected formation,
time slot reuse is not considered and the best superframe length is equal 
to the number of UAVs in the formation.

Simulations were run with $TSR=0$, $CT=7$, $ST=10$ and were repeated 
400 times. The parameters that resulted in the shortest time to convergence 
for the D-ART protocol were chosen. The final superframe size and 
convergence time (in rounds) for both the \gls{dst} and \gls{dar} protocol 
are captured in Table~\ref{tab:dst_dar_compare}.

The \gls{dst} protocol is shown to also work in the single-hop
scenario. This is indicated by the fact that the \gls{dst} protocol
achieves convergence, with the correct superframe size, across all the
simulations run.

The \gls{dst} protocol is also shown to outperform the \gls{dar}
protocol in nearly all cases. The time to convergence for \gls{dst} is
$17.1$ rounds shorter than \gls{dar} when the formation size and
starting superframe size are almost equal, and $73388.2$ rounds
shorter than \gls{dar} when the formation size is 100 times larger
than the starting superframe size. This further supports the analysis
that the growing mechanism proposed in the \gls{dst} protocol
alleviates contention in a small number of rounds.

The only case where the \gls{dst} protocol achieves convergence slower
than the \gls{dar} protocol is when the starting superframe size is 40
times larger than the formation size. This further supports the
analysis that the shrinking mechanism proposed in the \gls{dst}
protocol is the lengthiest part of reaching convergence.

\begin{table}[]
\centering
\begin{tabular}{|c|c|c|c|c|c|}
\cline{3-6}
\multicolumn{2}{c}{} & \multicolumn{2}{|c|}{Superframe Length} & \multicolumn{2}{|c|}{Convergence Time (rounds)} \\ \cline{1-6} 
U       & S         & \gls{dar}   & \gls{dst}  & \gls{dar}   & \gls{dst}        \\ \hline
200     & 200       & 200         & 200        & 38.7        & 21               \\ \hline
200     & 2         & 200         & 200        & 73441.5     & 53.3             \\ \hline
5       & 200       & 5           & 5          & 82          & 36543.3          \\ \hline
5       & 2         & 5           & 5          & 38.1        & 21               \\ \hline
\end{tabular}
\caption{Results comparing the \gls{dar} and the \gls{dst} protocol.}\label{tab:dst_dar_compare}
\end{table}

\gls{dst} reaches resolution at a comparable rate to previous work while being tested on a much larger number of nodes. 
In the literature, MD-MAC, Reins-MAC, and Z-MAC have been tested on randomly deployed networks of $20-200$ nodes with a 
maximum node degree $D=6$~\cite{zheng2019md}. In \cite{zheng2019md}, convergence is defined as the state where the 
transmission time slots for all nodes are spread with a spacing of at least $\frac{T}{n}$ and there are no collisions, 
where $T$ indicates the period and $n$ denotes the number of nodes in the network. Therefore, unlike in this work, 
convergence does not have the requirement of eliminating unused time slots. These protocols were shown to take 
$\sim17$ rounds for all devices to be successfully allocated to a time slot (i.e.\ to reach resolution). As shown in 
Table~\ref{tab:dst_avg_resolution}, \gls{dst} also takes $\sim17$ rounds to resolve when all nodes had degree $D=6$.

\begin{table}[]
	\centering
    \begin{tabular}{l|c|c}
	U     & Resolution Time \\\hline
	10    & 17.4			\\
	100   & 16.5 			 \\
	\end{tabular}
    \caption{The average normalized time to resolution and time to convergence (in rounds) at the upper and lower formation size settings.}
\label{tab:dst_avg_resolution}
\end{table}

\gls{dst} also achieves convergence (where there are no unused time slots) faster in networks with a higher 
node degree than MD-MAC, Reins-MAC, and Z-MAC achieve resolution. As seen in Table~\ref{tab:dst_dar_compare}, 
the average convergence times with \gls{dst} are $21$ or $53.3$ when the node degree is $D=200$ \glspl{u}. 
The average resolution times of MD-MAC, Reins-MAC, and Z-MAC with a node degree $D=16$ are $\sim60$ $\sim70$ 
and $\sim80$ rounds respectively.

\section{Conclusion}
\label{sec:conclusion}

The \gls{dst} protocol is shown to support distributed superframe
schedule creation and spatial reuse, achieving consensus and removing
unused time slots from the superframe.

A sensitivity study was conducted to determine the effect of the
protocol parameters on a number of outcomes of interest, namely time
to convergence, final superframe size, resolution time, overhead, and
the average number of \glspl{u} per time slot.  The protocol was shown
to be not sensitive to the size of the formation, with an increase in
the number of \glspl{u} only slightly increasing all outcomes of
interest on average, indicating that it is scalable.  The protocol
also maintains low overhead, achieves good reuse and always converges,
further supporting the \gls{dst} protocol's scalability.

The protocol was also tested against our centralized scheme to
determine how close the distributed protocol could come to this
benchmark.  In addition, the protocol was tested on a single-hop
formation against our distributed single-hop protocol to demonstrate
that the \gls{dst} protocol is suitable for both single-hop and
multi-hop scenarios.

Future work could include a comparison of these results with a CSMA
approach to quantify the difference for this use case, or for the 
use case of a dynamic multi-UAV system where the neighborhood of a
UAV changes throughout deployment. 

Our protocol assumes that a single UAV is `first' to join the formation.
Future work could also explore solutions where multiple UAVs believe they are
the first to join a formation and set up a superframe schedule concurrently.
In addition, protocol performance on formations with alternate structures 
or non-homogeneous node densities could be explored.


\bibliographystyle{elsarticle-num} 
\bibliography{main}

\end{document}